\documentclass{IEEEtran}
\usepackage{cite}
\usepackage{amsmath,amssymb,amsfonts}
\usepackage{algorithmic}
\usepackage{graphicx}
\usepackage{textcomp}
\usepackage{color,soul}
\usepackage{psfrag}
\usepackage{epstopdf}
\usepackage[mode=errorstop]{pstool}

\def\BibTeX{{\rm B\kern-.05em{\sc i\kern-.025em b}\kern-.08em
    T\kern-.1667em\lower.7ex\hbox{E}\kern-.125emX}}

\newcommand{\dya}[1]
{\overline{\overline{#1}}}

\begin{document}
\title{Comparison of \\ Tensor Boundary Conditions (TBCs) with \\ Generalized Sheet Transition Conditions~(GSTCs)}
\author{Mojtaba Dehmollaian, \IEEEmembership{Senior Member, IEEE}, Guillaume Lavigne, and Christophe Caloz, \IEEEmembership{Fellow, IEEE}
\thanks{The paper is submitted on ... 2019. \newline
M. Dehmollaian is with the Center of Excellence on Applied Electromagnetic Systems, School of Electrical and Computer Engineering, University of Tehran, 15719-14911, Tehran, Iran (e-mail: m.dehmollaian@ut.ac.ir). \newline
C. Caloz is with the Department of Electrical Engineering, Polytechnique Montr\'{e}al, H3T 1J4, Qu\'{e}bec, Canada (e-mail: guillaume.lavigne@polymtl.ca and christophe.caloz@polymtl.ca).}}

\maketitle

\begin{abstract}
This paper compares Tensor Boundary Conditions (TBCs), which were introduced to model multilayered dielectric structures, with Generalized Sheet Transition Conditions (GSTCs), which have been recently used to model metasurfaces. It shows that TBCs, with their 3 scalar parameters, are equivalent to the direct-isotropic -- cross-antiisotropic\footnote{What is meant here is the following. The term \emph{direct-isotropic} refers to scalar electric-to-electric and magnetic-to-magnetic susceptibilities, or direct susceptibilities. In contrast, the term \emph{cross-antiisotropic} refers to electric-to-magnetic and magnetic-to-electric susceptibilities, or cross-susceptibilities, that have opposite off-diagonal terms and zero diagonal terms. In other words, direct-isotropic refers to direct susceptibilities that are scalar in terms of the symmetric unit dyadic [Eq.~\eqref{eq:ed_iso_dir}], while cross-antiisotropic refers to cross susceptibilities that are scalar in terms of the antisymmetric unit dyadic [Eq.~\eqref{eq:ed_iso_cross}].}, reciprocal and nongyrotropic subset of GSTCs, whose 16~tangential (particular case of zero normal polarizations) or 36~general susceptibility parameters can handle the most general bianisotropic sheet structures. It further shows that extending that TBCs scalar parameters to tensors and allowing a doubly-occurring parameter to take different values leads to a TBCs formulation that is equivalent to the tangential GSTCs, but without reflecting the polarization physics of sheet media, such as metasurfaces and two-dimensional material allotropes.
\end{abstract}


\section{Introduction}
\label{sec:introduction}

\IEEEPARstart{R}ecently, the Generalized Sheet Transition Conditions (GSTCs) have been abundantly and successfully used as an electromagnetic model to synthesize and analyze metasurfaces~\cite{Achouri_Nanophotonics_2018,Vahabzadeh_JMMCT_2018}. GSTCs are generalizations of the conventional boundary conditions~\cite{Harrington_THEM_2001}, relating the field differences at both sides of a sheet discontinuity not only to surface currents but also to surface polarisations~\cite{Jia_TAP_2019} (Eqs.~(1)-(3) and Appendix~A therein). They were originally introduced by Idemen~\cite{Idemen_EL_1987,Idemen_DEF_2011}, next expressed in terms of surface polarizability or susceptibility tensors for metasurfaces by Kuester and Holloway~\cite{Kuester_TAP_2003,Holloway_APM_2012}, and finally extended to the most general\footnote{The only restriction of GSTCs as given in~\cite{Achouri_Nanophotonics_2018} in terms of generality is the fact that they are restricted to first-order discontinuities, i.e., to discontinuities that involve only the Dirac delta distribution and none of its derivatives, as the conventional boundary conditions. However, this is a rather minor restriction: this formulation seems to cover most of the practical sheet discontinuities.} case of bianisotropic (generally 36 parameters, and often 16 tangential parameters) metasurfaces by Achouri and Caloz~\cite{Achouri_Nanophotonics_2018,Achouri_TAP_2015}. GSTCs may also apply to two-dimensional materials constituted of a single layer or a small number of atomic layers, such as graphene, molybdenum disulfide or black phosphorous~\cite{Vahabzadeh_JMMCT_2018}.

Several extensions of the conventional boundary conditions have been reported in the literature~\cite{Senior_ABCE_1995}, and it is important to understand their similarities and differences with GSTCs for the optimal electromagnetic modeling of sheet structures. The greatest generality of the GSTCs compared to other boundary or transition conditions is most often obvious. However, there is an exception: the Tensor Boundary Conditions (TBCs) reported by Topsakal, Volakis and Ross in~\cite{Topsakal_RS_2002} may be written in a form that also involves a tensor (Eq.~(2) in~\cite{Topsakal_RS_2002}) despite their initial formulation in terms of just 3 scalar parameters~(Eq.~(1) in~\cite{Topsakal_RS_2002}). Given the quite different appearance of these TBCs compared to typical GSTCs (e.g. \cite{Achouri_Nanophotonics_2018}), the degree of similarity between the two classes of conditions is far from straightforward. Are the GSTCs just an alternative formulation of TBCs, or do they  indeed represent a more general description of sheet discontinuities, such as metasurfaces or two-dimensional material allotropes?

The present paper addresses this question. For this purpose, it expresses the TBCs in terms of equivalent surface tensorial susceptibilities, compares these susceptibilities with the GSTCs ones, and discusses the differences in terms of bianisotropy modeling.

\section{Mathematical Comparison}\label{sec:formulations}
\begin{figure}[h!t]
\centering
\psfragfig*[width=\columnwidth]{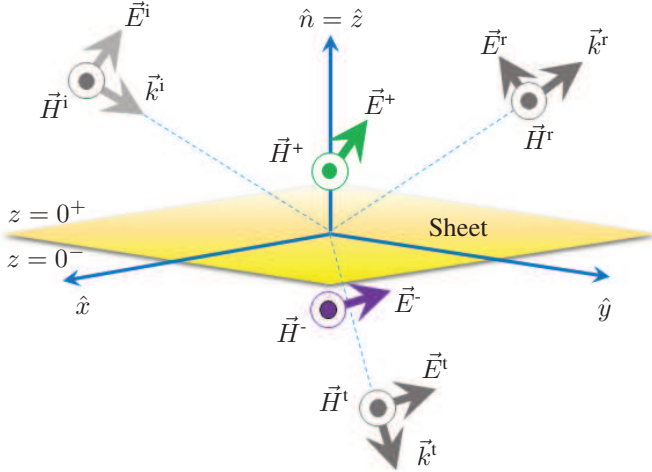}{
\psfrag{a}[][]{$\hat n=\hat z$}
\psfrag{b}[c][c]{$\vec E^{\text{i}}$}
\psfrag{c}[c][c]{$\vec H^{\text{i}}$}
\psfrag{d}[c][c]{$\vec k^{\text{i}}$}
\psfrag{e}[c][c]{$\vec E^{\text{r}}$}
\psfrag{f}[c][c]{$\vec H^{\text{r}}$}
\psfrag{g}[c][c]{$\vec k^{\text{r}}$}
\psfrag{h}[c][c]{$\vec E^{\text{t}}$}
\psfrag{i}[c][c]{$\vec H^{\text{t}}$}
\psfrag{j}[c][c]{$\vec k^{\text{t}}$}
\psfrag{k}[c][c]{$\vec E^{\text{+}}$}
\psfrag{l}[c][c]{$\vec H^{\text{+}}$}
\psfrag{m}[c][c]{$\vec E^{\text{-}}$}
\psfrag{n}[c][c]{$\vec H^{\text{-}}$}
\psfrag{o}[c][c]{$\text{Sheet}$}
\psfrag{x}[c][c]{$\hat x$}
\psfrag{y}[c][c]{$\hat y$}
\psfrag{p}[c][c]{$z=0^{+}$}
\psfrag{q}[c][c]{$z=0^{-}$}}
\caption{Problem of electromagnetic scattering from an anisotropic sheet of global (ee, me, em, mm) susceptibility $\dya{\chi}$. The sheet is located in the $xy$-plane, whose normal unit vector is $\hat n=\hat z$, at $z=0$. The electric and magnetic fields above and below the sheet are noted ($\vec E^+$, $\vec H^+$) and ($\vec E^-$, $\vec H^-$), respectively.}\label{fig:sheet_discont}
\end{figure}

For convenience, the parameters used in the paper are listed in Tab.~\ref{table_par}, and Fig.~\ref{fig:sheet_discont} depicts the problem of electromagnetic scattering from a bianisotropic sheet with relevant field quantities and coordinate system. The time convention $e^{-i \omega t}$ is implicitly assumed everywhere.
\begin{table}[h]
\caption{Parameters used in the paper.}
\label{table_par}
\centering
\setlength{\tabcolsep}{3pt}
\begin{tabular}{|p{25pt}|p{145pt}|p{18pt}|}
\hline
Symbol&
Quantity&
Unit \\
\hline
$c$&
speed of light in free-space&
m/s \\
$\omega$&
angular frequency&
rad/s \\
$k_0$&
free-space wavenumber&
rad/m \\
$\eta_0$&
free-space wave impedance ($\simeq 120 \pi$)&
$\Omega$ \\
$Z$&
surface impedance&
$\Omega$ \\
${\epsilon_0}$&
free-space permittivity ($\simeq 1/36 \pi \times 10^{-9}$)&
F/m \\
${\mu_0}$&
free-space permeability ($= 4 \pi \times 10^{-7}$)&
H/m \\
$\sigma$&
conductivity&
$\mho/$m \\
$\vec E$&
electric field&
V/m \\
$\vec H$&
magnetic field&
A/m \\
$\vec P$&
electric surface polarization density&
C/m \\
$\vec M $&
magnetic surface polarization density&
A \\
$\vec J_{\text{imp}\|}$&
impressed electric surface current density&
A/m \\
$\vec K_{\text{imp}\|}$&
impressed magnetic surface current density&
V/m \\
$\dya \chi_{\text{ee}\|}$&
electric-to-electric surface susceptibility&
m \\
$\dya \chi_{\text{em}\|}$&
magnetic-to-electric surface susceptibility&
m \\
$\dya \chi_{\text{mm}\|}$&
magnetic-to-magnetic surface susceptibility&
m \\
$\dya \chi_{\text{me}\|}$&
electric-to-magnetic surface susceptibility&
m \\
$R_{\text{e}}$&
scalar resistivity \cite{Topsakal_RS_2002}&
$\Omega$ \\
$R_{\text{m}}$&
scalar conductivity \cite{Topsakal_RS_2002}&
$\mho$ \\
$R_{\text{c}}$&
cross-coupling term \cite{Topsakal_RS_2002}&
1 \\
\hline
\end{tabular}
\end{table}

%\begin{figure}[t]
%\begin{center}
%\noindent
%  \includegraphics[width=\columnwidth]{sheet_discont.png}
%  \vspace{-4mm}
%  \caption{Problem of electromagnetic scattering from an anisotropic sheet of global (ee, me, em, mm) susceptibility $\dya{\chi}$. The sheet is located in the $xy$-plane, whose normal unit vector is $\hat n=\hat z$, at $z=0$. The electric and magnetic fields above and below the sheet are noted ($\vec E^+$, $\vec H^+$) and ($\vec E^-$, $\vec H^-$), respectively.}\label{fig:sheet_discont}
%\end{center}
%\end{figure}

\subsection{General Sheet Transition Conditions (GSTCs)}\label{subsec:GSTC}
The susceptibility-based GSTCs read~\cite{Achouri_Nanophotonics_2018,Jia_TAP_2019}
\begin{subequations}\label{eq:GSTC_DE}
\begin{equation}
\hat n \times \Delta \vec E = i\omega {\mu _0}{{\vec M}_\|}+\nabla_{\|}(P_{\bot}/\epsilon_0)-\vec{K}_{\text{imp}\|},
\end{equation}
\begin{equation}
\hat n \times \Delta \vec H =  -i\omega {{\vec P}_\|}-\hat n \times \nabla_{\|}M_{\bot}+\vec{J}_{\text{imp}\|},
\end{equation}
\end{subequations}
where $\hat n$ is the unit vector normal the surface of the sheet, the symbols $\|$ and $\bot$ denote vector components tangential and normal to the metasurface, respectively, and $\Delta$ refers to the difference of the fields at both sides of the metasurface, at $z=0^-$ and $z=0^+$. In the sequel, we assume, for simplicity, that\footnote{\label{fn:norm_pol} According to the Huygens principle, any physical fields on either side of the metasurface (i.e., everywhere outside a longitudinally ($z$) thin and transversally ($xy$) infinite volume containing the metasurface) can be produced by purely tangential equivalent surface polarizations. Therefore, a metasurface involving normal polarizations can always be reduced to an equivalent metasurface with purely tangential polarizations. However, the exclusion of normal components implies restrictions in terms of design flexibility and separate transformation possibility, which are discussed in~\cite{Achouri_Nanophotonics_2018}.} $P_{\bot}=M_{\bot}=0$, which reduces the coupled partial differential equations (\ref{eq:GSTC_DE}) to a simple system of algebraic linear equations.

The remaining tangential electric and magnetic surface polarizations in~\eqref{eq:GSTC_DE} relate to the averaged fields as
\begin{subequations}\label{MP}
\begin{align}\label{MPa}
{\vec M}_\|=
\begin{pmatrix}
{{M_x}}\\
{{M_y}}
\end{pmatrix}
& = \frac{1}{\eta_0}
\begin{pmatrix}
{\chi _\text{me}^{xx}}&{\chi _\text{me}^{xy}}\\
{\chi _\text{me}^{yx}}&{\chi _\text{me}^{yy}}
\end{pmatrix}
\begin{pmatrix}
{E_{x,\,\text{av}}}\\
{E_{y,\,\text{av}}}
\end{pmatrix} \nonumber \\
& +
\begin{pmatrix}
{\chi _\text{mm}^{xx}}&{\chi _\text{mm}^{xy}}\\
{\chi _\text{mm}^{yx}}&{\chi _\text{mm}^{yy}}
\end{pmatrix}
\begin{pmatrix}
{H_{x,\,\text{av}}}\\
{H_{y,\,\text{av}}}
\end{pmatrix} \nonumber \\
& ={\frac{1}{\eta_0}} \overline {\overline {\chi}} _{\text{me}\|} \cdot \vec E_{\|,\,\text{av}} + \overline{\overline {\chi}} _{\text{mm}\|} \cdot \vec H_{\|,\,\text{av}},
\end{align}
\begin{equation}
{{\vec P}_\|} =
\begin{pmatrix}
{{P_x}}\\
{{P_y}}
\end{pmatrix}
= {\epsilon _0} \overline {\overline {\chi}} _{\text{ee}\|} \cdot \vec E_{\|,\,\text{av}} + \frac{1}{c} \overline{\overline {\chi}} _{\text{em}\|} \cdot \vec H_{\|,\,\text{av}},
\end{equation}
\end{subequations}
where ``av'' refers to the average of the fields at both sides of the metasurface at $z=0^-$ and $z=0^+$, and ${\dya{\chi}}_{\text{me}\|}$, ${\dya{\chi}}_{\text{mm}\|}$, ${\dya{\chi}}_{\text{ee}\|}$ and ${\dya{\chi}}_{\text{em}\|}$ are the magnetic-to-electric, magnetic-to-magnetic, electric-to-electric and magnetic-to-electric surface susceptibility tensors, respectively~\cite{Achouri_Nanophotonics_2018}.

Inserting Eqs.~\eqref{MP} into Eqs.~\eqref{eq:GSTC_DE} with $P_{\bot}=M_{\bot}=0$ while assuming that no impressed surface current densities are present ($\vec{J}_{\text{imp}\|}=\vec{K}_{\text{imp}\|}=0$) yields
\begin{subequations}\label{GSTC}
\begin{equation}
\hat z \times \Delta \vec E
=ik_0{\dya {\chi}} _{\text{me}\|} \cdot \vec E_{\|,\,\text{av}}
+ik_0 {\eta _0}{\dya{\chi}} _{\text{mm}\|} \cdot \vec H_{\|,\,\text{av}},
\end{equation}
\begin{equation}
\hat z \times \Delta \vec H = - ik_0/{\eta _0}{\dya{\chi}} _{\text{ee}\|} \cdot \vec E_{\|,\,\text{av}}-ik_0{\dya {\chi}} _{\text{em}\|} \cdot \vec H_{\|,\,\text{av}},
\end{equation}
\end{subequations}
which express the difference fields in terms of the average fields via the susceptibility tensors. The GSTCs (\ref{GSTC}), characterized by their 4 tangential susceptibility tensors (16 scalar parameters\footnote{The assumption $P_{\bot}=M_{\bot}=0$ (or $P_{\bot}$ and $M_{\bot}$ space-wise constant) in~\eqref{eq:GSTC_DE} indeed reduces the effect of the most general bianisotropic $3\times 3$ ($4\times(3\times 3)=36$ scalar parameters) susceptibility tensors to that of their $2\times 2$ ($4\times(2\times 2)=16$ scalar parameters) tangential parts. If the bianisotropic sheet includes normal polarizations (e.g. in-plane rings, leading to nonzero $M_\perp$), the normal susceptibility components $\chi_{ab}^{xz}$, $\chi_{ab}^{yz}$, $\chi_{ab}^{zx}$, $\chi_{ab}^{zy}$,  and $\chi_{ab}^{zz}$ ($a,b=\text{e,m}$) may also  be involved.}), can essentially model any transverse-bianisotropic metasurface that supports first-order induced polarization surface currents~\cite{Idemen_DEF_2011,Achouri_TAP_2015}.

\subsection{Tensor Boundary Conditions (TBCs)}\label{subsec:TBC}
The TBCs, as given by Eq.~(1) in~\cite{Topsakal_RS_2002}, read
\begin{subequations}\label{eq:TBC_O}
\begin{align}
\hat n \times &\left( \vec E^{+} + \vec E^{-}  \right) = R_{\text{e}}~\hat n \times \left(\hat n \times \vec H^{+} - \hat n \times \vec H^{-} \right) \nonumber \\
&-R_{\text{c}}~\hat n \times \left( \hat n \times \left( \hat n \times \vec E^{+} - \hat n \times \vec E^{-} \right) \right) ,
\end{align}
\begin{align}
\hat n \times &\left( \vec H^{+} + \vec H^{-}  \right) = R_{\text{m}}~\hat n \times \left(\hat n \times \vec E^{+} -\hat n \times \vec E^{-} \right)  \nonumber \\
&+R_{\text{c}}~\hat n \times \left( \hat n \times \left( \hat n \times \vec H^{+} - \hat n \times \vec H^{-} \right) \right),
\end{align}
\end{subequations}
where $\vec E^{\pm}$ and $\vec H^{\pm}$ are the electric and magnetic fields at $z=0^\pm$. Using the field average and difference notation in Sec.~\ref{sec:formulations} (e.g., $\vec E_{\text{av}}=(\vec E^{+} + \vec E^{-})/2$ and $\Delta \vec E=(\vec E^{+} - \vec E^{-})$) and setting $\hat n = \hat z$, these equations take the more compact form
\begin{subequations}\label{eq:TBC}
\begin{equation}
2\hat z \times \vec E_{\|,\,\text{av}} = R_{\text{e}}~\hat z \times \left(\hat z \times \Delta \vec H \right) + R_{\text{c}}~\hat z \times \Delta \vec E,
\end{equation}
\begin{equation}
2\hat z \times \vec H_{\|,\,\text{av}} = R_{\text{m}}~\hat z \times \left(\hat z \times \Delta \vec E \right) - R_{\text{c}}~\hat z \times \Delta \vec H.
\end{equation}
\end{subequations}
These equations relate the average fields to the difference fields via the 3 scalar parameters $R_{\text{e}}$, $R_{\text{m}}$, and $R_{\text{c}}$, whose units are respectively $\Omega$, $\mho$ and $1$.

\subsection{GSTCs Susceptibilities in Terms of TBCs Parameters}\label{subsec:TBC}

The TBCs, as given by Eqs.~\eqref{eq:TBC_O} or~\eqref{eq:TBC}, have an obvious disadvantage compared to susceptibility-based GSTCs for handling complex sheets, such as metasurfaces or magnetized two-dimensional materials: They are not directly expressed in terms of bianisotropic medium parameters, and therefore provide little insight into the physics of the problem.

However, solving~\eqref{eq:TBC} for the difference fields leads to the following reformulation in terms of TBCs-equivalent susceptibilities  (see Appendix~A):
\begin{subequations}\label{eq:TBC_to_GSTC}
\begin{equation}\label{eq:TBC_to_GSTC_a}
\hat z \times \Delta \vec E
=ik_0\dya{\chi}_{\text{me}\|}^\text{TBC}\cdot\vec E_{\|,\,\text{av}}
+ik_0\eta_0\dya{\chi}_{\text{mm}\|}^\text{TBC}\cdot\vec H_{\|,\,\text{av}},
\end{equation}
\begin{equation}\label{eq:TBC_to_GSTC_b}
\hat z \times \Delta \vec H
=-ik_0/\eta_0~\dya{\chi}_{\text{ee}\|}^\text{TBC}\cdot\vec E_{\|,\,\text{av}}
-ik_0\dya{\chi}_{\text{em}\|}^\text{TBC}\cdot
\vec H_{\|,\,\text{av}},
\end{equation}
where
\begin{equation}\label{eq:TBC_to_GSTC_c}
\dya{\chi}_{\text{me}\|}^\text{TBC}
=i\frac{2R_{\text{c}}}{k_0D}
\begin{pmatrix}
0 & -1 \\
1 & 0
\end{pmatrix},\quad
\dya{\chi}_{\text{mm}\|}^\text{TBC}
=-i\frac{2R_{\text{e}}}{k_0\eta_0D}
\begin{pmatrix}
1 & 0 \\
0 & 1
\end{pmatrix},
\end{equation}
\begin{equation}\label{eq:TBC_to_GSTC_d}
\dya{\chi}_{\text{ee}\|}^\text{TBC}
=i\frac{2\eta_0R_{\text{m}}}{k_0D}
\begin{pmatrix}
1 & 0 \\
0 & 1
\end{pmatrix},\quad
\dya{\chi}_{\text{em}\|}^\text{TBC}
=i\frac{2R_{\text{c}}}{k_0D}
\begin{pmatrix}
0 & -1 \\
1 & 0
\end{pmatrix},
\end{equation}
with
\begin{equation}
D=R_{\text{e}}R_{\text{m}}-R_{\text{c}}^2.
\end{equation}
\end{subequations}
%
%\begin{subequations}\label{eq:TBC_to_GSTC}
%\begin{equation}
%\hat z \times \Delta \vec E
%=\dya{\chi}_{\text{me}\|}^\text{TBC}\cdot\vec E_{\|,\,\text{av}}
%+\dya{\chi}_{\text{mm}\|}^\text{TBC}\cdot\vec H_{\|,\,\text{av}},
%\end{equation}
%\begin{equation}
%\hat z \times \Delta \vec H
%=\dya{\chi}_{\text{ee}\|}^\text{TBC}\cdot\vec E_{\|,\,\text{av}}
%+\dya{\chi}_{\text{em}\|}^\text{TBC}\cdot
%\vec H_{\|,\,\text{av}},
%\end{equation}
%where
%\begin{equation}\label{eq:TBC_to_GSTC_c}
%\dya{\chi}_{\text{me}\|}^\text{TBC}
%=-\frac{2R_{\text{c}}}{D}
%\begin{pmatrix}
%0 & -1 \\
%1 & 0
%\end{pmatrix},\quad
%\dya{\chi}_{\text{mm}\|}^\text{TBC}
%=\frac{2R_{\text{e}}}{D}
%\begin{pmatrix}
%1 & 0 \\
%0 & 1
%\end{pmatrix},
%\end{equation}
%\begin{equation}\label{eq:TBC_to_GSTC_d}
%\dya{\chi}_{\text{ee}\|}^\text{TBC}
%=\frac{2R_{\text{m}}}{D}
%\begin{pmatrix}
%1 & 0 \\
%0 & 1
%\end{pmatrix},\quad
%\dya{\chi}_{\text{em}\|}^\text{TBC}
%=\frac{2R_{\text{c}}}{D}
%\begin{pmatrix}
%0 & -1 \\
%1 & 0
%\end{pmatrix},
%\end{equation}
%with
%\begin{equation}
%D=R_{\text{e}}R_{\text{m}}-R_{\text{c}}^2.
%\end{equation}
%\end{subequations}

Comparing the partial transverse susceptibilities in Eqs.~\eqref{eq:TBC_to_GSTC} with the full ones in Eqs.~\eqref{GSTC} reveals that the TBCs in~\cite{Topsakal_RS_2002} form a restricted subset of the TBCs, with the following 14~restrictions:
\begin{subequations}\label{eq:tens_restr}
\begin{equation}\label{eq:tr_ee}
\chi_\text{ee}^{xy}=\chi_\text{ee}^{yx}=0,
\end{equation}
\begin{equation}\label{eq:tr_mm}
\chi_\text{mm}^{xy}=\chi_\text{mm}^{yx}=0,
\end{equation}
\begin{equation}\label{eq:tr_em}
\chi_\text{em}^{xx}=\chi_\text{em}^{yy}=0,
\end{equation}
\begin{equation}\label{eq:tr_me}
\chi_\text{me}^{xx}=\chi_\text{me}^{yy}=0,
\end{equation}
\end{subequations}

and

\begin{subequations}\label{eq:tr_self}
\begin{equation}\label{eq:tr_self_a}
\chi_\text{ee}^{xx}=\chi_\text{ee}^{yy},
\end{equation}
\begin{equation}\label{eq:tr_self_b}
\chi_\text{mm}^{xx}=\chi_\text{mm}^{yy},
\end{equation}
\begin{equation}\label{eq:tr_self_c}
\chi_\text{em}^{xy}=-\chi_\text{em}^{yx},
\end{equation}
\begin{equation}\label{eq:tr_self_d}
\chi_\text{me}^{xy}=-\chi_\text{me}^{yx},
\end{equation}
\end{subequations}

and

\begin{subequations}\label{eq:tr_cross}
\begin{equation}
\chi_\text{me}^{xy}=-\chi_\text{em}^{yx},
\end{equation}
\begin{equation}
\chi_\text{me}^{xy}=-\chi_\text{em}^{yx}.
\end{equation}
\end{subequations}

\section{Physical Restrictions of the TBCs}\label{sec:properties}

The tensorial conditions~\eqref{eq:tens_restr} are associated to restrictions on the transverse bianisotropic physical properties of the sheet.

\subsection{Direct Isotropy -- Cross Antiisotropy}
Equations~\eqref{eq:tr_ee}-\eqref{eq:tr_self_a}, and \eqref{eq:tr_mm}-\eqref{eq:tr_self_b} respectively imply that
\begin{subequations}\label{eq:ed_iso_dir}
\begin{equation}\label{eq:ed_iso_dir_a}
\dya{\chi}_{\text{ee}\|}=\chi_{\text{ee}\|}\dya{I}_{\|}^{\text{S}},
\end{equation}
\begin{equation}\label{eq:ed_iso_dir_b}
\dya{\chi}_{\text{mm}\|}=\chi_{\text{mm}\|}\dya{I}_{\|}^{\text{S}},
\end{equation}
\end{subequations}
where $\dya{I}_{\|}^{\text{S}}=\hat{x}\hat{x}+\hat{y}\hat{y}$ is the symmetric transverse unit dyadic tensor, while Eqs.~\eqref{eq:tr_em}-\eqref{eq:tr_self_c} and \eqref{eq:tr_me}-\eqref{eq:tr_self_d} respectively imply that
\begin{subequations}\label{eq:ed_iso_cross}
\begin{equation}
\dya{\chi}_{\text{em}\|}=\chi_{\text{em}\|}\dya{I}_{\|}^{\text{A}},
\end{equation}
\begin{equation}
\dya{\chi}_{\text{me}\|}=\chi_{\text{me}\|}\dya{I}_{\|}^{\text{A}},
\end{equation}
\end{subequations}
where $\dya{I}_{\|}^{\text{A}}=\hat{y}\hat{x}-\hat{x}\hat{y}$ is the antisymmetric transverse unit dyadic tensor.

Equations~\eqref{eq:ed_iso_dir} indicate that the electric-to-electric and magnetic-to-magnetic susceptibilities are scalar (electric and magnetic isotropy), while~\eqref{eq:ed_iso_cross} indicate that the magnetic-to-electric and electric-to-magnetic susceptibilities are tensorial with zero diagonal terms and opposite off-diagonal terms, or ``antiisotropic.''

According to the conventional definition of biisotropy~\cite{Kong_EMT_2008,Lindell_EWCBM_1994} -- all the constitutive parameters scalar -- such a sheet would not be (transversally) biisotropic. However, one may argue that antiisotropy, as defined by~\eqref{eq:ed_iso_cross}, is closer to special form of isotropy than to anisotropy. Indeed, rotating the coordinate system about the $\hat{z}$-axis (i.e., substituting $x\rightarrow y$ and $y\rightarrow-x$) in the susceptibilities~\eqref{eq:TBC_to_GSTC_c}-\eqref{eq:TBC_to_GSTC_d} leaves the system~\eqref{eq:TBC_to_GSTC_a}-\eqref{eq:TBC_to_GSTC_d} unchanged, which means that the sheet has properties that are transversally isotropic, assuming uniformity\footnote{Consider for instance a periodically nonuniform metasurface with the same periodicity (meant here as the \emph{medium} periodicity corresponding to a supercell periodicity, not to the sampling periodicity of the unit-cell discretization) along the $x$- and $y$- directions (with periods $L_x$ and $L_y$). Such a metasurface can obviously not be isotropic, since the period in the $xy$-direction ($\sqrt{L_{\smash{x}}^{\smash{2}}+L_{\smash{y}}^{\smash{2}}}$) is different from that in the $x$- and $y$- directions}, just as a conventionally-called isotropic structure.

Thus, the TBCs in~\cite{Topsakal_RS_2002} are restricted to direct-isotropic cross-antiisotropic sheets. This turns out to correspond to the form of bianisotropy that is required for diffractionless generalized refraction~\cite{Lavigne_TAP_2018} or for any power-conserving nongyrotropic transformation~\cite{Epstein_TAP_2016}, except for the restriction to single (p or s) polarization\footnote{In nongyrotropic bianisotropic metasurfaces, the $p$ and $s$ polarizations are decoupled. For instance, for a plane wave incident in the $xz$ plane, we have separate $p$~($E_x$ and $H_y$ transverse fields) and $s$~($E_y$ and $H_x$ transverse fields) polarizations. The scattering for the $p$ polarization is controlled by the susceptibility components $\chi_\text{ee}^{{xx}}$, $\chi_\text{mm}^{{yy}}$, $\chi_\text{em}^{{xy}}$, and $\chi_\text{me}^{{yx}}$, and the scattering for the $s$ polarization can be independently controlled by $\chi_\text{ee}^{{yy}}$, $\chi_\text{mm}^{{xx}}$, $\chi_\text{em}^{{yx}}$, and $\chi_\text{me}^{{xy}}$. Therefore, the GSTCs, having all the susceptibility degrees of freedom, can handle these polarizations separately. In contrast, in the TBCs, the responses for the two polarizations are dependent from each other via the restrictions in Eqs.~\eqref{eq:tr_self}. Hence, TBCs cannot handle the $p$ and $s$ polarizations separately: once a design has been made for one, the design for the other one is constrained by it, which dramatically restricts the available transformations. For example, the TBCs cannot model a metasurface supporting refraction for both polarizations, as shown in Appendix~\ref{sec:refraction}.}. However, this specific form of anisotropy naturally involves restrictions, such as those that will be described in the next two subsections.

\subsection{Reciprocity}

The condition for nonreciprocity are~\cite{Achouri_Nanophotonics_2018,Kong_EMT_2008,Caloz_PRA_2018}
\begin{subequations}\label{eq:NR_cond}
\begin{equation}\label{eq:NR_cond_a}
{\dya{\chi}}_{\text{ee}\|}\neq{\dya{\chi}}_{\text{ee}\|}^T
\end{equation}
or
\begin{equation}\label{eq:NR_cond_b}
{\dya{\chi}}_{\text{mm}\|}\neq{\dya{\chi}}_{\text{mm}\|}^T
\end{equation}
or
\begin{equation}\label{eq:NR_cond_cd}
{\dya{\chi}}_{\text{me}\|}\neq-{\dya{\chi}}_{\text{em}\|}^T.
\end{equation}
\end{subequations}
where the symbol $T$ denotes the transpose operation.

According to~\eqref{eq:ed_iso_dir_a} and~\eqref{eq:ed_iso_dir_b} (resp.), the conditions~\eqref{eq:NR_cond_a} and~\eqref{eq:NR_cond_b} (resp.) cannot be satisfied. Moreover, Eqs.~\eqref{eq:tr_cross} prohibit the satisfaction of Eq.~\eqref{eq:NR_cond_cd}.

Thus, the TBCs in~\cite{Topsakal_RS_2002} are restricted, from Eqs.~\eqref{eq:tens_restr}, \eqref{eq:tr_self} and~\eqref{eq:tr_cross} (14 conditions), to reciprocal~\cite{Achouri_Nanophotonics_2018,Kong_EMT_2008} sheets. For instance, they cannot handle spatial-isolation metasurfaces~\cite{Kodera_AWPL_2018}.

%If the medium is lossless, then transpose operation in Eqs.~\eqref{eq:NR_cond} must be replaced by the transpose conjugate operation, which implies for nonreciprocity that $\text{Im}\{\dya{\chi}_\text{ee}\}\neq 0$ or $\text{Im}\{\dya{\chi}_\text{mm}\}\neq 0$ or $\text{Re}\{\dya{\chi}_\text{em}\}\neq 0$ or
%$\text{Re}\{\dya{\chi}_\text{em}\}\neq 0$~\cite{Caloz_PRA_2018}. This more restrictive (lossless) form of nonreciprocity

\subsection{Nongyrotropy}

The condition for gyrotropy is~\cite{Achouri_Nanophotonics_2018,Kong_EMT_2008}
\begin{subequations}\label{eq:gyr_cond}
\begin{equation}\label{eq:gyr_cond_a}
\chi_{\text{ee}}^{xy},\chi_{\text{ee}}^{yx}\neq 0
\end{equation}
or
\begin{equation}\label{eq:gyr_cond_b}
\chi_{\text{mm}}^{xy},\chi_{\text{mm}}^{yx}\neq 0
\end{equation}
or
\begin{equation}\label{eq:gyr_cond_cd}
\chi_{\text{em}}^{xx},\chi_{\text{em}}^{xx},
\chi_{\text{me}}^{xx},\chi_{\text{me}}^{yy}
\neq 0.
\end{equation}
\end{subequations}

All of these relations are prohibited by~\eqref{eq:tens_restr} (8 conditions). Thus, the TBCs in~\cite{Topsakal_RS_2002} are restricted to be nongyrotropic\footnote{In a nongyrotropic medium, the transverse electric (TE) and transverse magnetic (TM) polarizations are always decoupled.}. For instance, they cannot handle polarization rotators~\cite{Grbic_PRA_2014}.

\section{Tensorial Extension of the TBCs}\label{sec:derivation_tensor}

We have found in Secs.~\ref{sec:formulations} and~\ref{sec:properties} that the TBCs in~\cite{Topsakal_RS_2002} represent a subset of the GSTCs with restriction to direct-isotropic/cross-antiistropic susceptibilities. However, one may then legitimately ask whether making the scalar parameters in the TBCs~\eqref{eq:TBC} tensorial would provide the same level of generality as the GSTCs. The present section addresses this question.

Transforming the scalar parameters $R_\text{e}$, $R_\text{m}$ and $R_\text{c}$ into tensors, and allowing the third parameter to take different values at its two occurrences for maximal freedom reformulates~\eqref{eq:TBC} as
\begin{subequations}\label{eq:TBC_T}
\begin{equation}
2\hat z\times \vec E_{\|,\,\text{av}}=\dya R_{\text{e}}\cdot \left(\hat z\times \left(\hat z\times \Delta \vec H \right)\right)+\dya R_{\text{ce}}\cdot \left(\hat z\times \Delta \vec E \right),
\end{equation}
\begin{equation}
2\hat z\times \vec H_{\|,\,\text{av}}=\dya R_{\text{m}}\cdot \left(\hat z\times \left(\hat z\times \Delta \vec E \right)\right)-\dya R_{\text{cm}}\cdot \left(\hat z\times \Delta \vec H \right),
\end{equation}
\end{subequations}

In order to derive the susceptibilities associated with Eq.~(\ref{eq:TBC_T}), one first needs, as for the scalar-parameter case (Sec.~\ref{sec:derivation}), to express the difference fields in terms of the average fields for proper comparison with GSTCs. For this purpose, we vectorially pre-multiply both sides of Eqs.~\eqref{eq:TBC_T} by $\hat{z}$, which is easily achieved after noticing that the operator $\hat{z}\times\vec{v}_\|$ for any transverse vector $\vec{v}_\|$ is equivalent to the operator $\dya{N}\cdot\vec{v}_\|$, where
\begin{equation}
\dya N
=\begin{pmatrix}
{0}&{-1}\\
{1}&{0}
\end{pmatrix}
\end{equation}
The result is
\begin{subequations}\label{eq:TBC_TC}
\begin{equation}
-2\vec E_{\|,\,\text{av}} = \dya N\cdot\dya R_{\text{e}}\cdot\dya N\cdot\left(\hat z \times \Delta \vec H \right) + \dya N\cdot\dya R_{\text{ce}}\cdot\left(\hat z \times \Delta \vec E\right),
\end{equation}
\begin{equation}
-2\vec H_{\|,\,\text{av}} = \dya N\cdot\dya R_{\text{m}}\cdot\dya N\cdot\left(\hat z \times \Delta \vec E \right) - \dya N\cdot\dya R_{\text{cm}}\cdot\left(\hat z \times \Delta \vec H\right).
\end{equation}
\end{subequations}

Solving these equations for the vectors $\hat z \times \Delta \vec E$ and $\hat z \times \Delta \vec H$ yields
\begin{subequations}\label{eq:TBC_sol}
\begin{align}
\hat z \times \Delta \vec E = &-2\left(\dya N\cdot\dya R_\text{e}\cdot\dya N\cdot\dya D_\text{e} \right)^{-1}\cdot\vec E_{\|,\,\text{av}} \nonumber \\ &-2\left(\dya  N\cdot\dya R_\text{cm}\cdot\dya D_\text{e} \right)^{-1}\cdot\vec H_{\|,\,\text{av}},
\end{align}
\begin{align}
\hat z \times \Delta \vec H = &-2\left(\dya  N\cdot\dya R_\text{ce}\cdot\dya D_\text{m} \right)^{-1} \vec E_{\|,\,\text{av}} \nonumber \\ &+ 2\left(\dya N\cdot\dya R_\text{m}\cdot\dya N\cdot\dya D_\text{m} \right)^{-1}\cdot\vec H_{\|,\,\text{av}},
\end{align}
where
\begin{equation}
\dya D_\text{e}=\dya R_\text{cm}^{-1}\cdot\dya R_\text{m}\cdot\dya N-\dya N\cdot\dya R_\text{e}^{-1}\cdot\dya R_\text{ce},
\end{equation}
\begin{equation}
\dya D_\text{m}=\dya R_\text{ce}^{-1}\cdot\dya R_\text{e}\cdot\dya N-\dya N\cdot\dya R_\text{m}^{-1}\cdot\dya R_\text{cm}.
\end{equation}
\end{subequations}

Comparing Eqs.~(\ref{eq:TBC_sol}) and (\ref{GSTC}) provides then the TBCs-equivalent susceptibility tensors
\begin{subequations}\label{eq:susc_eqv}
\begin{equation}
\dya{\chi}_{\text{me}\|}^\text{TBC}=i\frac{2}{k_0}\left(\dya N\cdot\dya R_\text{e}\cdot\dya N\cdot\dya D_\text{e} \right)^{-1},
\end{equation}
\begin{equation}
\dya{\chi}_{\text{mm}\|}^\text{TBC}=i\frac{2}{k_0\eta_0}\left(\dya N\cdot\dya R_\text{cm}\cdot\dya D_\text{e} \right)^{-1},
\end{equation}
\begin{equation}
\dya{\chi}_{\text{ee}\|}^\text{TBC}=-i\frac{2\eta_0}{k_0}\left(\dya N\cdot\dya R_\text{ce}\cdot\dya D_\text{m} \right)^{-1},
\end{equation}
\begin{equation}
\dya{\chi}_{\text{em}\|}^\text{TBC}=i\frac{2}{k_0}\left(\dya N\cdot\dya R_\text{m}\cdot\dya N\cdot\dya D_\text{m} \right)^{-1},
\end{equation}
\end{subequations}
which may easily be verified to reduce to Eqs.~\eqref{eq:TBC} upon reducing the tensors $\dya{R}$ reduce to tensors and setting $R_\text{ce}=R_\text{cm}=R_\text{c}$. Since these tensors may be completely different from each other, the susceptibilities~\eqref{eq:susc_eqv} involve $4\times(2\times 2)=16$ independent parameters, and Eqs.~\eqref{eq:TBC_T} are therefore perfectly equivalent to the GSTCs~\eqref{GSTC}.

However, two important comments are here in order. First, the expressions~\eqref{GSTC} are overly complicated, and fail to properly represent the polarization physics~\cite{Jackson_CE_1998} involved in the sheet media. Second, in contrast to the most general GSTCs~\eqref{eq:GSTC_DE}, which may involve $4\times(3\times 3)=36$ surface susceptibility parameters through their spatial derivatives (Footnote~\ref{fn:norm_pol}), the TBCs~\eqref{eq:TBC} are always restricted to $4\times(2\times 2)=16$ susceptibility parameters.

\section{Conclusion}\label{sec:con}
TBCs, as given in~\cite{Topsakal_RS_2002}, are not equivalent to GSTCs, as given in~\cite{Achouri_Nanophotonics_2018}. They represent a subset of GSTCs which can only
model  sheets that are direct-isotropic -- cross-antiisotopic, reciprocal and nongyrotropic. Moreover, they are restricted to 16 equivalent susceptibility parameters, whereas GSTCs, in their most general form, support the 36 susceptibility parameters corresponding to the three dimensions of space. Finally, in contrast to GSTCs, they do not reflect the polarization physics of sheet media such as metasurfaces and two-dimensional material allotropes.
\appendices
\numberwithin{figure}{section}
\numberwithin{equation}{section}
\section{Derivation of the GSTCs Susceptibilities Corresponding to the TBCs}\label{sec:derivation}

The TBCs equations~\eqref{eq:TBC} may be recast in the matrix form
\begin{subequations}\label{eq:TBC_M}
\begin{equation}
2\begin{pmatrix}
-E_{y,\text{av}}\\
E_{x,\text{av}}
\end{pmatrix}
=R_{\text{e}}
\begin{pmatrix}
-\Delta H_{x}\\
-\Delta H_{y}
\end{pmatrix}
+R_{\text{c}}
\begin{pmatrix}
-\Delta E_{y}\\
\Delta E_{x}
\end{pmatrix},
\end{equation}
\begin{equation}
2\begin{pmatrix}
-H_{y,\text{av}}\\
H_{x,\text{av}}
\end{pmatrix}
=R_{\text{m}}
\begin{pmatrix}
-\Delta E_{x}\\
-\Delta E_{y}
\end{pmatrix}
-R_{\text{c}}
\begin{pmatrix}
-\Delta H_{y}\\
\Delta H_{x}
\end{pmatrix}.
\end{equation}
\end{subequations}
This system of equations may be rearranged in terms of the electromagnetically uncoupled equation pairs
\begin{subequations}
\begin{equation}\label{eq:TBC_set1}
\begin{split}
-2 E_{y,\text{av}} = -R_{\text{e}}\Delta H_{x}-R_{\text{c}}\Delta E_{y},\\
2 H_{x,\text{av}} = -R_{\text{m}}\Delta E_{y}-R_{\text{c}}\Delta H_{x},
\end{split}
\end{equation}
and
\begin{equation}\label{eq:TBC_set2}
\begin{split}
2 E_{x,\text{av}} = -R_{\text{e}}\Delta H_{y}+R_{\text{c}}\Delta E_{x},\\
-2 H_{y,\text{av}} = -R_{\text{m}}\Delta E_{x}+R_{\text{c}}\Delta H_{y}.
\end{split}
\end{equation}
\end{subequations}
Solving Eq.~(\ref{eq:TBC_set1}) for $\Delta H_{x}$ and $\Delta E_{y}$ yields
\begin{subequations}\label{GSTC_set1}
\begin{equation}
\Delta H_{x} = \frac{2R_{\text{m}}}{D}E_{y,\text{av}}+\frac{2R_{\text{c}}}{D}H_{x,\text{av}},
\end{equation}
\begin{equation}
\Delta E_{y} = \frac{-2R_{\text{c}}}{D}E_{y,\text{av}}-\frac{2R_{\text{e}}}{D}H_{x,\text{av}},
\end{equation}
\end{subequations}
with
\begin{equation}
D=R_{\text{e}}R_{\text{m}}-R_{\text{c}}^2.
\end{equation}
Similarly solving Eq.~(\ref{eq:TBC_set2}) for $\Delta H_{y}$ and $\Delta E_{x}$ yields
\begin{subequations}\label{GSTC_set2}
\begin{equation}
\Delta H_{y} = \frac{-2R_{\text{m}}}{D}E_{x,\text{av}}+\frac{2R_{\text{c}}}{D}H_{y,\text{av}},
\end{equation}
\begin{equation}
\Delta E_{x} = \frac{-2R_{\text{c}}}{D}E_{x,\text{av}}+\frac{2R_{\text{e}}}{D}H_{y,\text{av}}.
\end{equation}
\end{subequations}
Rearranging Eqs.~(\ref{GSTC_set1}) and (\ref{GSTC_set2}) in the matrix form
\begin{subequations}\label{eq:GSTC_M}
\begin{equation}
\begin{split}
\begin{pmatrix}
\Delta E_{y}\\
-\Delta E_{x}
\end{pmatrix}
=
&\frac{-2R_{\text{e}}}{D}
\begin{pmatrix}
H_{x,\text{av}}\\
H_{y,\text{av}}
\end{pmatrix}\\
&+ \frac{2R_{\text{c}}}{D} \begin{pmatrix}
0&-1\\
1&0
\end{pmatrix}
\begin{pmatrix}
E_{x,\text{av}}\\
E_{y,\text{av}}
\end{pmatrix},
\end{split}
\end{equation}
\begin{equation}
\begin{split}
\begin{pmatrix}
-\Delta H_{y}\\
\Delta H_{x}
\end{pmatrix}
=
&\frac{2R_{\text{m}}}{D}
\begin{pmatrix}
E_{x,\text{av}}\\
E_{y,\text{av}}
\end{pmatrix}\\
&+\frac{2R_{\text{c}}}{D}
\begin{pmatrix}
0&-1\\
1&0
\end{pmatrix}
\begin{pmatrix}
H_{x,\text{av}}\\
H_{y,\text{av}}
\end{pmatrix},
\end{split}
\end{equation}
\end{subequations}
leads to the GSTCs-form equations [Eqs.~\eqref{eq:TBC_to_GSTC}]
\begin{subequations}\label{eq:GSTC_F}
\begin{equation}
\hat z \times \Delta \vec E = \frac{2R_{\text{e}}}{D} \vec H_{\|,\,\text{av}} - \frac{2R_{\text{c}}}{D}
\begin{pmatrix}
0&-1\\
1&0
\end{pmatrix}
\vec E_{\|,\,\text{av}},
\end{equation}
\begin{equation}
\hat z \times \Delta \vec H = \frac{2R_{\text{m}}}{D} \vec E_{\|,\,\text{av}} + \frac{2R_{\text{c}}}{D}
\begin{pmatrix}
0&-1\\
1&0
\end{pmatrix}
\vec H_{\|,\,\text{av}}.
\end{equation}
\end{subequations}

\section{Generalized Refraction Metasurface Example}\label{sec:refraction}

Consider a metasurface surrounded by air on both sides that is designed to refract a plane wave incident at an angle $\theta_\text{i}$ towards an angle
$\theta_\text{t}$. The susceptibility functions  characterizing such a metasurface for the $p$ polarization, given in~\cite{Lavigne_TAP_2018}, can be written as

\begin{subequations}\label{eq:p_refraction}
\begin{equation}\label{eq:p_refraction_xee}
\chi_\text{ee}^{xx}=\frac{2 T (4 \sin \alpha  \cos \theta_\text{i} +\beta  T \sin (2 \alpha ))}{\eta_0 \omega  \epsilon_0  (2 \cos \theta_\text{i} +\beta  T \cos \alpha )^2},
\end{equation}
\begin{equation}\label{eq:p_refraction_xmm}
\chi_\text{mm}^{yy}=\frac{2 \eta_0 T \cos \theta_\text{i} \cos \theta_\text{t} (4 \sin \alpha  \cos \theta_\text{i}+\beta  T \sin (2 \alpha ))}{\mu_0  \omega  (2 \cos \theta_\text{i}+\beta  T \cos \alpha )^2},
\end{equation}

\begin{equation}\label{eq:p_refraction_xem}
\chi_\text{me}^{yx}= -\chi_\text{em}^{xy} = \frac{2 (T \cos \alpha  (\cos \theta_\text{i}-\cos \theta_\text{t}))}{k_0 (2 \cos \theta_\text{i}+\beta  T \cos \alpha )},
\end{equation}
\end{subequations}
\noindent where $\alpha = k_0 x (\sin \theta_\text{i} - \sin \theta_\text{t})$, $\beta = \cos \theta_\text{i} +\cos \theta_\text{t}$ and
 $T = \sqrt{\cos \theta_\text{i} / \cos \theta_\text{t}}$. The susceptibility functions for the $s$ polarization are found by duality as
\begin{subequations}\label{eq:s_refraction}
\begin{equation}\label{eq:s_refraction_xee}
\chi_\text{ee}^{yy}=\frac{2 T \cos \theta_\text{i} \cos \theta_\text{t} (4 \sin \alpha  \cos \theta_\text{i}+\beta  T \sin (2 \alpha ))}{\eta_0 \omega  \epsilon_0  (2 \cos \theta_\text{i}+\beta  T \cos \alpha )^2},
\end{equation}
\begin{equation}\label{eq:s_refraction_xmm}
\chi_\text{mm}^{xx}= \frac{2 \eta_0 T (4 \sin \alpha  \cos \theta_\text{i}+\beta  T \sin (2 \alpha ))}{\mu_0  \omega  (2 \cos \theta_\text{i}+\beta  T \cos \alpha )^2},
\end{equation}

\begin{equation}\label{eq:s_refraction_xem}
\chi_\text{me}^{xy}= -\chi_\text{em}^{yx} = -\frac{2 (T \cos \alpha  (\cos \theta_\text{i}-\cos \theta_\text{t}))}{k_0 (2 \cos \theta_\text{i}+\beta  T \cos \alpha )}.
\end{equation}
\end{subequations}

Comparing Eqs~\eqref{eq:p_refraction} and~\eqref{eq:s_refraction} shows that this set of susceptibility functions violates the restrictions of
Eqs.~\eqref{eq:tr_self} of the TBCs. Hence, such a metasurface cannot be described using by the TBCs as presented in~\cite{Topsakal_RS_2002}.


\begin{thebibliography}{00}

\bibitem{Achouri_Nanophotonics_2018}
    K. Achouri and C. Caloz, ``Design, concepts, and applications of electromagnetic metasurfaces,'' \emph{Nanophotonics}, vol.~7, no.~6, pp.~1095–-1116, Jun.~2018.

\bibitem{Vahabzadeh_JMMCT_2018}
    Y. Vahabzadeh, N. Chamanara, K. Achouri, and C. Caloz, ``Computational analysis of metasurfaces,'' \emph{J. Multiscale Multiphys. Comput. Tech.}, vol.~3, pp.~37–-49, Apr.~2018.

\bibitem{Harrington_THEM_2001} R. F. Harrington, \emph{Time-Harmonic Electromagnetic Fields}, Hoboken, USA, Wiley-IEEE Press, 2nd ed., 2001.

\bibitem{Jia_TAP_2019} X. Jia, Y. Vahabzaeh, C. Caloz, and F. Yang, ``Synthesis  of Spherical Metasurfaces based on Susceptibility Tensor GSTCs,'' \emph{IEEE Trans. Antennas Propag.}, {\it to be published}.

\bibitem{Idemen_EL_1987}
    M. Idemen A. Hamit and Serbest, ``Boundary conditions of the electromagnetic field,'' \emph{Electron. Lett.}, vol.~23, no.~13, pp.~704--705, Jun.~1987.

\bibitem{Idemen_DEF_2011} M. M. Idemen, \emph{Discontinuities in the Electromagnetic Field}, Hoboken, IET, Wiley, 2011.

\bibitem{Kuester_TAP_2003} E. F. Kuester, M. A. Mohamed, M. Piket-May, and C. L. Holloway, ``Averaged transition conditions for electromagnetic fields at a metafilm,''  \emph{IEEE Trans. Antennas Propag.}, vol.~51, no.~10, pp.~2641--2651, Oct.~2003.

\bibitem{Holloway_APM_2012} C. L. Holloway, E. F. Kuester, J. A. Gordon, J. O. Hara, J. Booth, and D. R. Smith, ``An overview of the theory and applications of metasurfaces: The two-dimensional equivalents of metamaterials,'' \emph{IEEE Antennas Propag. Mag.}, vol.~54, no.~2, pp.~10–-35, Apr.~2012.

 \bibitem{Achouri_TAP_2015} K. Achouri, M. A. Salem, and C. Caloz, ``General metasurface synthesis based on susceptibility tensors,'' \emph{IEEE Trans. Antennas Propag.}, vol.~63, no.~7, pp.~2977–-2991, Jul.~2015.

\bibitem{Senior_ABCE_1995} T. B. A. Senior and J. L. Volakis, \emph{Approximate Boundary Conditions in Electromagnetics}, IET,~1995.

\bibitem{Topsakal_RS_2002} E. Topsakal, J. L. Volakis, and D. C. Ross, ``Surface integral equations for material layers modeled with tensor boundary conditions,'' \emph{Radio Science}, vol.~37, no.~4, pp.~1-6, Jul.~2002.

\bibitem{Kong_EMT_2008} J. A. Kong, \emph{Electromagnetic Wave Theory}, Cambridge, USA, 2008.

\bibitem{Lindell_EWCBM_1994} I. V. Lindell, A. H. Sihvola, S. A. Tretyakov, and A. J. Viitanen, \emph{Electromagnetic Mixing Frmulas and Applications}, Artech House, 1999.

\bibitem{Lavigne_TAP_2018} G. Lavigne, K. Achouri, V. S. Asadchy, S. A. Tretyakov, and C. Caloz, ``Susceptibility derivation and experimental demonstration of refracting metasurfaces without spurious diffraction,'' \emph{IEEE Trans. Antennas Propag.}, vol.~66, no.~3, pp.~1321–-1330, Mar.~2018.

\bibitem{Epstein_TAP_2016} A. Epstein and G. V. Eleftheriades, ``Arbitrary power-conserving field transformations with passive lossless omega-type bianisotropic metasurfaces,'' \emph{IEEE Trans. Antennas Propag.}, vol.~64, no.~9, pp.~3880--3895, Sep.~2016.

\bibitem{Caloz_PRA_2018} C. Caloz, A. Al\`{u}, S. Tretyakov, D. Sounas, K. Achouri, and Z.-L. Deck-Léger, ``Electromagnetic nonreciprocity,'' Phys. Rev. Appl., vol.~10, no.~4, pp.~047 001:1-–26, Oct.~2018.

\bibitem{Kodera_AWPL_2018} T. Kodera and C. Caloz, S. Tretyakov, D. Sounas, K. Achouri, and Z.-L. Deck-Léger, ``Unidirectional loop metamaterials ({ULM}) as magnetless artificial ferrimagnetic materials: principles and applications,'' IEEE Antennas Wirel. Propag. Lett., vol.~17, no.~11, pp.~1943–-1947, Nov.~2018.

\bibitem{Grbic_PRA_2014} C. Pfeiffer and A. Grbic, ``Bianisotropic metasurfaces for optimal polarization control: analysis and synthesis,'' Phys. Rev. Appl., vol.~2, no.~044011, pp.~1–-11, Feb.~2014.

\bibitem{Jackson_CE_1998} J. D. Jackson, \emph{Classical Electrodynamics}, Wiley, third.~ed., 1998.

\end{thebibliography}
\end{document}